\documentclass[prl,twocolumn,amsmath,amssymb,superscriptaddress,showpacs] {revtex4-2}

\usepackage{mathbbol,amsmath,amsfonts,amssymb,bm,xcolor,graphicx,psfrag,array,mathtools,lipsum,mathrsfs,comment,adjustbox,stmaryrd,braket,cancel,soul}
\usepackage{subfig}
\usepackage{rotate,fancybox,epic,amsbsy,epsfig,graphics}
\usepackage[justification=Justified]{caption}

\usepackage{hyperref}
\usepackage{graphicx,psfrag,color}
\usepackage{dcolumn}
\usepackage{bm}
\usepackage{mathbbol, appendix}
\usepackage{printlen}
\usepackage[justification=Justified]{caption}
\usepackage{diagbox}
\usepackage{comment}
\usepackage[normalem]{ulem}
\usepackage{float}

\newcolumntype{t}[1]{D{.}{.}{#1}}

\usepackage{natbib}
\begin{document}

\title{Topological Orders Beyond
Topological Quantum Field Theories}

\author{P. Vojta}
\affiliation{Department of Physics, Washington University in St. Louis, MO 63130 USA}
\author{G. Ortiz}
\affiliation{Department of Physics, Indiana University, Bloomington, IN 47405, USA}
\affiliation{Institute for Advanced Study, Princeton, NJ 08540, USA}
\affiliation{Institute for Quantum Computing, University of Waterloo, Waterloo, N2L 3G1, ON, Canada}
\author{Z. Nussinov}
\email{corresponding author: zohar@wustl.edu}
\affiliation{Rudolf Peierls Centre for Theoretical Physics, University of Oxford, Oxford OX1 3PU, United Kingdom}
\affiliation{Department of Physics, Washington University in St. Louis, MO 63130 USA}
\affiliation{Institute for Theoretical Solid State Physics, IFW Dresden, Helmholtzstrasse 20, 01069 Dresden, Germany}

\date{\today}

\begin{abstract}
Systems displaying quantum topological order feature robust characteristics that are very attractive to quantum computing schemes. Topological quantum field theories have proven to be powerful in capturing the quintessential attributes of systems displaying  topological order including, in particular, their anyon excitations. Here, we investigate systems that lie outside this common purview, and present a rich class of models exhibiting topological orders with distance-dependent interactions between anyons. As we illustrate, in some instances, {\it the gapped lowest-energy excitations are comprised of anyons that densely cover the entire system}. This leads to behaviors not typically described by topological quantum field theories. We examine these models by performing exact dualities to systems displaying conventional (i.e., Landau) orders. Our approach enables a general method for mapping generic Landau-type theories to dual models with topological order of the same spatial dimension. The low-energy subspaces of our models can be made more resilient to thermal effects than those of surface codes.
\end{abstract}

\maketitle

{\it Introduction.}
Topological quantum field theories (TQFTs)\cite{Witten:1988ze,Atiyah_1988,WittenQFTJones,lurie2009classification}, are closely interwoven with existing descriptions of topological quantum order (TQO). Axiomatically, TQFTs are mappings from (inherently 
metric independent) manifold  cobordisms to Hilbert spaces \cite{Atiyah_1988}. It is often understood that a microscopic model displaying TQO effectively renormalizes to a certain TQFT in the low-energy limit, resulting in ground-state degeneracies computable via that TQFT \cite{gukov2013topological,Chang:2018iay}.    
Consequently, many distinct microscopic models may be associated with the same type of TQO. Importantly, from a practical standpoint, the properties of {\it  gapped low-energy (anyon) excitations} in TQO systems are typically analyzed 
via such 
TQFTs \cite{Simon}.These excitations form the focus of our attention. TQFTs are indispensable in elegantly studying ground states and viable basic  charges (anyons) in the lowest energy excited states of general TQO theories (including anyon braiding and fusion) \cite{Simon}.
With the exception of theories exhibiting  rigid subsystem symmetries (e.g., fractons and related models), where the degeneracy increases with system size, \cite{Haah,Yoshida_2013,fracton1,Slagle_2018,fracton2,Weinstein_2019,fracton3,exp_deg}, TQFTs generally encode ground-state  (anyon vacuum) degeneracies. We will show that excitations of TQO systems may differ from those of expected anyon models. The interactions in these systems can non-trivially alter the braiding and fusion properties of their lowest-energy excitations  (these excitations may become unbounded so as to cover all of space) without modifying the ground-state space TQFT. Even in the infrared limit, 
whenever
(anyon) excitations appear,
their 
salient characteristics will be metric (i.e., geometry) dependent. TQFT descriptions of the vacuum 
remain unaltered. 
To elucidate the basic premise, we introduce and study simple 
$\mathbb{Z}_q$ (with $q\ge 2$) extensions of the 
well-known ($q=2$) 
Kitaev toric code (TC)  \cite{kitaev_2003} models. 
Our construct is not limited to these models and also applies to string-net   \cite{SM,Lin_2021_genStringNet,Levin_2005, Buerschaper_2009,Pachos_2012} long-range,  specific non-abelian \cite{bondalgebras}, high-dimensional, and other specific extensions of the TC \cite{SM}.
As we illustrate, such TQO models may map, via exact dualities, to theories exhibiting Landau orders. 

Diverse, and often inequivalent \cite{nussinov_ortiz_2009_IsingChains,PNAS-top}, notions of TQO abound. We follow Kitaev's definition 
which highlights the innate robustness of TQO systems to (quasi-) local perturbations \cite{kitaev_2003}, hence their promise for topological quantum information processing. According to this error-detection motivated definition,  
the matrix elements of all (quasi-)local operators $V$ in the ground-state basis, spanned by orthonormal states $\{|{\sf g}_{\gamma} \rangle\}_{\gamma=1,N_{\sf g}}$, satisfy 
\begin{equation}
\label{KitaevCond}\bra{{\sf g}_{\alpha}}V\ket{{\sf g}_{\beta}} = v \delta_{\alpha,\beta} ,
\end{equation}
 with $v$ a constant depending only on $V$. Physically, Eq. (\ref{KitaevCond}) asserts that 
different ground-states cannot be told apart via (quasi-) local measurements (and thus cannot be assigned different conventional Landau order parameters). Such a condition may be extended beyond the lowest-energy (i.e., ground-state) sector  \cite{nussinov_ortiz_2009_IsingChains,PNAS-top} and finite temperature \cite{Nussinov_2008_3DTC}.
For symmetry enriched topological order (SET) \cite{Mesaros_2013}, $v$ may include quantum numbers associated with symmetries. 

\begin{figure*}[htb] 
\hspace*{0.0cm}\includegraphics[trim=0 0.75cm 0 0,height=36ex]{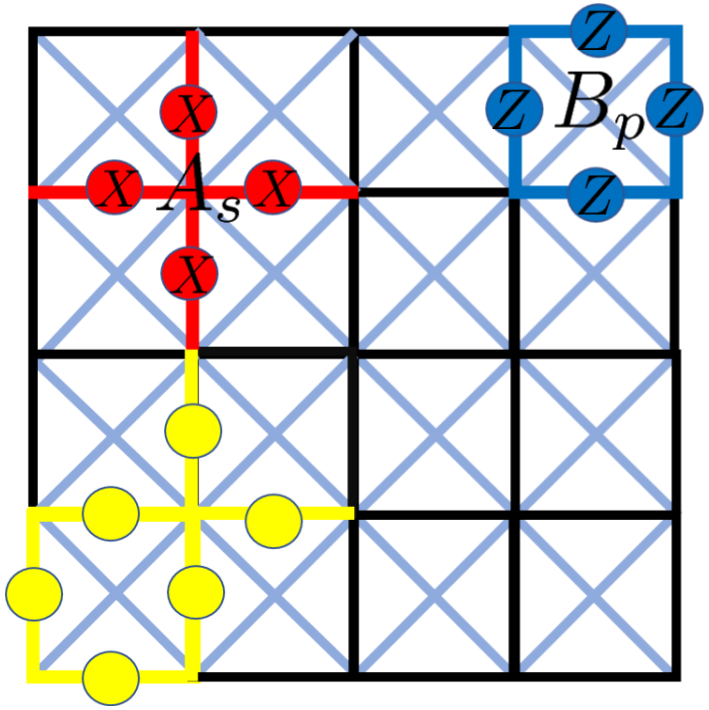}
\hspace*{1.0cm}
\includegraphics[trim=0 0.5 0 0.5cm,height=35ex]{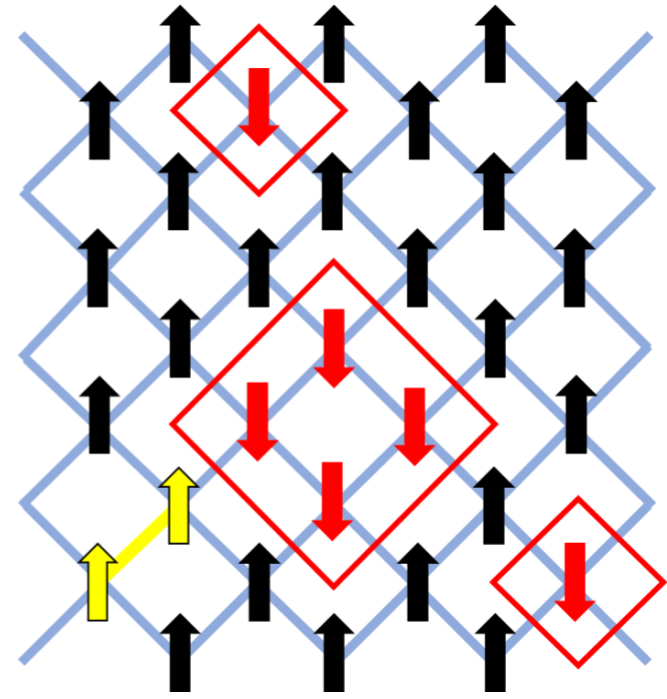}
\hspace*{1.4cm}\includegraphics[trim = 0 0 0 0, height=35ex]{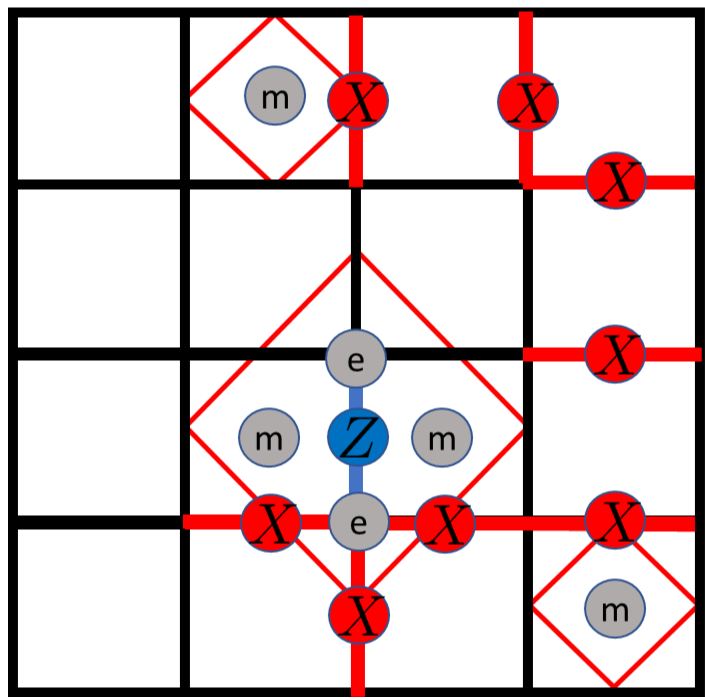}
  \caption{(Color online.) The  \(\mathbb{Z}_2\) SPPM as an example of our duality. Left: The SPPM of Eq. (\ref{SPPMHamiltonian}) with star $A_s$, plaquette $B_p$, and interaction $A_sB_p$ (yellow) operators indicated. Center: The associated dual classical Ising (\(q\) = 2) model. The interactions $A_sB_p$ are dual to the nearest-neighbor Ising spin interactions (also highlighted in yellow). Right: An excited SPPM state created by acting with
  string operations on the ground-state (the dual excited Ising model state is displayed in the central panel). $X$ and $Z$ string excitations create \(\rm m\) and \(\rm e\) anyons at their endpoints.}  
 \label{LatticeMapFig}
\end{figure*}

{\it The Star-Plaquette Product Model {\rm (SPPM)}.}
We highlight the families of Hamiltonians arising in  stabilizer codes. Our simplest non-trivial example will be the ``$q$-state SPPM.'' 
On 
lattice $\Lambda$, this model is given by the nearest-neighbor Hamiltonian (see Fig. \ref{LatticeMapFig}),
\begin{equation}\label{SPPMHamiltonian}
H_{\sf SPPM} = -\sum_{\langle s,p \rangle } J_{sp} A_{s}B_{p}-\sum_{s}g_{s} A_{s}-\sum_{p} g_{p} B_{p} +h.c. ,
\end{equation}
defined on a Hilbert space ${\cal H}$ formed  by the tensor product of $N_\Lambda$ link subspaces. The $\langle s, p \rangle$ sum in the first term is over neighboring ``stars'' and ``plaquettes'' (see left panel of Fig. \ref{LatticeMapFig}) sharing common lattice links. We will analyze excitations in systems having spatially  non-uniform couplings yet largely consider uniform $g_{s} = g_{p}=g$ and $J_{sp}=J$. In the square lattice realization of Eq. (\ref{SPPMHamiltonian}), the individual (commuting) star and plaquette operators are \(A_s =X_1X_2^\dagger X^\dagger_3 X_4\) and \(B_p = Z_1Z_2Z^\dagger_3 Z^\dagger _4\) with the labelling  performed anti-clockwise from the upper link of the star/plaquette, and where $\dagger$ is the adjoint. The elementary  $\mathbb{Z}_q$ ``clock'' and ``shift''  unitary operators $Z_j$ and $X_j$ lie on each of the $N_{\Lambda}$ lattice links $j$ and satisfy the Weyl group relation
\(X_j Z_j = e^{i 2\pi/q}Z_j X_j\) 
\cite{Ortiz_2012} (generalizing the two-state ($q=2$) Pauli group). The union of sites ($s$) and plaquette centers ($p$) or dual-sites forms a ``diagonal lattice''  highlighted in blue in Fig. \ref{LatticeMapFig} (a.k.a., ``radial lattice/graph'' \cite{abrams2019basic}). For $J,g>0$, the ground-states of the SPPM and the paradigmatic \(\mathbb{Z}_q\) TC model  \cite{kitaev_2003,Schulz_2012,gottesman1997stabilizer,Araujo_de_Resende_2019}, 
\begin{eqnarray}
\label{ZqTC}
H_{\sf TC} =  -\sum_{s}(A_s+A^\dagger_s)-\sum_{p}(B_p+B^\dagger _p),
\end{eqnarray}
are identical with a degeneracy exponential in the genus of the lattice surface 
\(N_{{\bf g},{\sf TC}} = q^{2({\sf genus})}\). 
The SPPM and TC also share all of their excited eigenstates (with degeneracies that are integer multiples of $N_{{\bf g},{\sf TC}}$). Crucially, however, the respective energies of these eigenstates will differ in both models. The first term in  Eq. (\ref{SPPMHamiltonian}) does not appear in the TC and leads to an energy that depends on the geometrical arrangement of TC defects (anyons) where ${\sf Eigenvalue}[A_{s}], {\sf Eigenvalue}[B_{p}] \neq 1$. 

For lattices $\Lambda$ on a torus, the minimal degeneracy of each energy level of both the SPPM and the TC is associated with a pair of ($d=1$ dimensional) non-commuting symmetries (for each independent non-contractible loop $C_{a}$ (or $C'_{a}$)) given by closed string products of (i) \(Z_j\) operators on the lattice $\Lambda$ and (ii) of \(X_j\) operators on the dual lattice $\Lambda_{\sf d}$,
\begin{equation}
\label{SymmetryLoops}
Z^{\sf q}_{1(2)}=\prod_{j \in C_{1(2)}} Z_j , \quad
X^{\sf q}_{1(2)}=\!\!\prod_{j \in C'_{1(2)}} \!\! X_j.
\end{equation}
The lowest--energy excited states of the TC model are created by acting on its ground-states with products of local \(Z_j\) or 
local \(X_j\) operators along open strings \cite{kitaev_2003}. We denote, respectively, the lattice and dual lattice strings by \(Z^{\sf q}_{s_1,s_2}\) and \(X^{\sf q}_{p_1,p_2}\) for strings with endpoints at 
sites \(s_1,s_2\) or dual-sites \(p_1,p_2\). Each such string product creates an excited state of fixed energy, regardless of its length, with anyons at its endpoints. That is, there is no ``string tension.'' The TC anyons can drift apart an arbitrary distance without energy cost. This feature makes only the topology of the anyons (not their  geometry) germane. The SPPM of \eqref{SPPMHamiltonian} fundamentally differs from the TC by having additional nearest-neighbor interactions appearing in the first term. These 
interactions 
produce differences in energies between nearest-neighbor and more distant anyons, introducing an effective string tension. The associated energies are identical to those of domain walls in classical clock models.





{\it Duality between the {\rm SPPM} and classical clock models.}
We next demonstrate that the SPPM is dual to a 
nearest-neighbor $q$-state 
clock model. 
For concreteness, we analyze the SPPM on a torus. The ``bond-algebra'' \cite{bondalgebras,nussinov_ortiz_2009_IsingChains,Ortiz_2012,BO-solns,BO-short,BO-Majorana,BA-orbital}, i.e., the set of all independent algebraic relations amongst the individual terms appearing in the SPPM Hamiltonian is given by 
\begin{subequations}
\begin{align}
[A_{s}, A_{s'}]=[A_{s}, B_{p}]=[B_{p},B_{p'}]&=& 0,
\label{SPPMConstra}
\\ 
A^q_s = B^q_ p &=& 1, 
\label{SPPMConstrb}
\\  
 (A_{s_1}B_{p_1})(A^\dagger_{s_2}B^\dagger_{p_1})(A_{s_2}B_{p_2})\cdots(A_{s_n}B_{p_n}) \nonumber \\
 \times (A^\dagger_{s_1}B^\dagger_{p_n}) &=& 1,
\label{SPPMConstrc} 
\\
 \prod_sA_s = \prod_pB_p &=& 1. 
\label{SPPMConstrd}
\end{align}
\end{subequations}
Eqs. (\ref{SPPMConstrc}) applies to the {\it product} of said bilinears {\it along any closed loop} $\Gamma$ of length $(2n)$. Eqs. (\ref{SPPMConstrc}) are identities that follow, algebraically, from Eqs. (\ref{SPPMConstra}). We have explicitly written down Eqs. (\ref{SPPMConstrc}) in order to  provide all bond-algebraic relations (including constraints amongst the individual bonds (terms) that appear in the SPPM Hamiltonian of Eq. (\ref{SPPMHamiltonian})).
Relations similar to Eqs. (\ref{SPPMConstrc}) apply to products of the $(A_s B_p)$ bilinears (and their Hermitian conjugates) along an {\it arbitrary open contour}  $\Gamma'$; when multiplied at their two endpoints ($r$ and $m$) by terms of the $A^{\dagger}_r$ and/or $B_m^{\dagger}$ type (or their Hermitian conjugates), such operator products along any  $\Gamma'$ are unity. %
Eqs. (\ref{SPPMConstra}, \ref{SPPMConstrb}, \ref{SPPMConstrc}) describe the bulk theory. By contrast, Eqs. (\ref{SPPMConstrd}) 
arise from periodic boundary conditions. Here, in a full lattice tiling with all stars $s$ or plaquettes $p$, in the products of Eqs. (\ref{SPPMConstrd}), each local unitary $X$-type operator or $Z$-type operator will appear twice.

We now turn to {\it classical}  Landau-type nearest-neighbor $q$-state clock models (classical 2D Ising model for $q=2$) in ``external longitudinal  fields" $g_i$ given by the Hamiltonian 
\begin{equation}
\label{HCLOCK}
H_{\sf c} = -\sum_{\langle i,j\rangle}J_{ij} (z^*_{i}z_{j}+z_{i}z^*_{j})-\sum_{i} g_{i} (z_{i} +z^*_i),
\end{equation}
with \(z_{j} \in \mathbb{C}\)
being \(q\)-th roots of unity.  
$H_{\sf c}$ is defined on the diagonal lattice of the SPPM, see Fig. 
\ref{LatticeMapFig}. 
The diagonal lattice splinters into ``even'' and ``odd" sub-lattices that are, respectively,  associated with the sites and dual-sites of the SPPM square lattice; 
by comparison to Eq. (\ref{SPPMHamiltonian}), we set $g_i = g_s$ (even sub-lattice) and $g_{i} = g_{p}$ (odd sub-lattice) with $J_{ij} = J_{sp}$. Aside from boundary consequences (Eq. (\ref{SPPMConstrd})), the duality between the SPPM and the clock model is a mapping of star operators to even-site clock variables and plaquettes to odd-site clock variables and vice versa, 
\begin{eqnarray}
\label{map1}
A_s \leftrightarrow  z_i, ~ ~B_p \leftrightarrow z^*_j.
\end{eqnarray}
This bond-algebraic duality mapping generates the terms in the clock model from those of the SPPM, the nearest-neighbor SPPM products map to nearest-neighbor products of the classical clock model $A_sB_p \leftrightarrow z_iz^*_j$. 
Apart from trivial (classical)  commutativity of its individual terms (similar to that of the quantum SPPM (Eq. (\ref{SPPMConstra})), the bond-algebra of $H_{\sf c}$ is specified by counterparts to Eqs. (\ref{SPPMConstrb}, \ref{SPPMConstrc}),
 \begin{subequations}
\begin{align}
z_i^q = 1, ~ \label{ClockConstraintb}   \\
\prod_{(i,j) \in \Gamma}z_i^*z_j = 1. 
\label{ClockConstrd}
\end{align}
\end{subequations}
Further analogous to the SPPM, the product $z_{r'} \Big(\prod_{(i,j) \in \Gamma'}z_i^*z_j \Big) z^*_{m'}= 1$ along {\it any open contour} $\Gamma'$ having $r'$ and $m'$ as its endpoints. 
Similar to Eqs. (\ref{SPPMConstrc}), these last  open contour product relations and the closed contour ($\Gamma$) product equalities of Eqs. (\ref{ClockConstrd}) are algebraic identities. 
An analog of Eq. (\ref{SPPMConstrd}) for the classical Ising ($q=2$) case would imply that all spin flips relative to the ferromagnetic state in the dual classical Ising model can only appear in pairs on any of the sublattices; for the general $q$-state clock model of Eq. (\ref{HCLOCK}), on each of the two  sublattices, the sum of the clock spin angles must be an integer multiple of $2 \pi$. A constraint equivalent to Eq. (\ref{SPPMConstrd}) can be realized in a clock model Hamiltonian dual 
 by amending $H_{\sf c}$ by a pair of non-local terms \cite{SM}. In the thermodynamic limit, 
the free energy density of the classical clock model  
is unaltered and the duality of Eq. (\ref{map1}) follows. 
Dualities between classical clock spin and SPPM systems for other boundary conditions 
can be found in \cite{SM}. Our duality implies that adding the $A_s B_p$ coupling term to the TC Hamiltonian $H_{\sf TC}$ , yielding $H_{\sf SPPM}$, modifies the classical one-dimensional clock (and Ising) model spectrum of $H_{\sf TC}$ \cite{Nussinov_2008_3DTC,nussinov_ortiz_2009_IsingChains,PNAS-top} to be that of the classical two-dimensional $H_{\sf c}$. 

From  Eq. (\ref{map1}), properties of the SPPM model can be inferred from well-known behaviors of classical clock (and Ising) models. We list several of these. When $g=0^{+}$, the global $\mathbb{Z}_q$ symmetry 
of the uniform coupling classical clock model is spontaneously broken at low temperatures with {\it free energy barriers  
diverging with system size}. When $g=0$, the $q>4$ clock models exhibit Kosterlitz-Thouless (KT) transitions (with two dual KT phase endpoints \cite{Ortiz_2012}); critical Ising and three-state Potts transitions appear for $q=2,4$ and $q=3$ renditions respectively. $H_{\sf SPPM}$ realizations with random $J_{sp}$ are dual to classical spin glass \cite{SG1} models. When both $J$ and $g$ are uniform non-zero constants, instead of a phase transition (i.e., a non-analyticity of the free energy in the thermodynamic limit), the system may exhibit a Widom line or a weak, Kertesz \cite{kertesz} type, crossover known in spin models and lattice gauge theories \cite{PRD} and thus similarly do so in its SPPM dual. For random $g_{s,p}$, the SPPM maps into a classical random field model featuring numerous effects  \cite{RF}. Classical ANNNI models arising for next-nearest-neighbor couplings $J_{ij}$ \cite{ANNNI} exhibit rich devil staircase structures. From the bond-algebraic type \cite{bondalgebras,nussinov_ortiz_2009_IsingChains,Ortiz_2012,BO-solns,BO-short,BO-Majorana,BA-orbital} duality (\ref{map1}), all such behaviors 
of these and other classical clock type systems appear in their quantum duals. Our construct further  demonstrates  that there is no  single fixed point associated with anyons of a given type. For instance, the duality of Eq. (\ref{map1}) transforms the Ashkin Teller model  \cite{aoun2023phase_AshkinTeller} (displaying a line of fixed points) to a topological system with TC-type anyons \cite{SM} displaying also a line of fixed points.  


{\it Topological Order of the {\rm SPPM}}.  
The duality of Eq. (\ref{map1}) implies that any 
eigenstate 
of the clock model is also an eigenstate of the SPPM. Particularly, when $J,g >0$, including the $g=0^{+}$ limit, 
the ground-states of the SPPM are eigenstates of all $A_s$ and $B_p$ operators with an eigenvalue of 1 \footnote{At low temperatures, the $g=0$ SPPM (dual to a two-dimensional clock model with no applied field) spontaneously breaks global two-dimensional $\mathbb{Z}_q$ symmetry. When this global symmetry is broken,  
Eq. (\ref{KitaevCond}) is violated when $V$ is set equal to the pertinent local order parameters (the SPPM $A_s$ or $B_p$) which, e.g., assume $q$  different expectation values $\langle A_s \rangle = \langle B_p \rangle$ in the disparate ground-state sectors (with each of 
these sectors being of dimension $q^{2({\sf genus})}$).}. 
By virtue of the symmetries  (\ref{SymmetryLoops}), the TC model satisfies the condition of Eq. (\ref{KitaevCond}) and its finite temperature extension \cite{nussinov_ortiz_2009_IsingChains,PNAS-top}, and thus harbors TQO. Similar to the TC model, these (low-dimensional) generalized gauge-like symmetries \cite{nussinov_ortiz_2009_IsingChains,PNAS-top,generalized,Elitzur1} cannot be spontaneously broken  \cite{Elitzur1,Elitzur2,Nussinov_2008_3DTC} and endow the system with TQO \cite{nussinov_ortiz_2009_IsingChains,PNAS-top}. Since 
the eigenstates of the SPPM are those of the TC model, 
it follows that the SPPM exhibits TQO. Equivalently, any given classical clock state $\{z_i\}$  
corresponds to a $N_{{\bf g},{\sf TC}}-$dimensional sector of topologically degenerate SPPM eigenstates of identical $\{A_{s},B_{p}\}$ eigenvalues that thus cannot be told apart by local measurements \footnote{Operators having different expectation values involve the non-local symmetries (\ref{SymmetryLoops}).}. The latter {\it topology dependent} degeneracy follows from our duality by an 
exponentiation, $q^{{\sf E}-{\sf F}-{\sf V}}= q^{2({\sf genus}-1)}$, of the Euler-Poincare formula to count redundant $q$-state operators. This is the number by which the $N_{\Lambda}$ local SPPM $\mathbb{Z}_q$ operators (number of lattice edges $\sf E$) exceeds the number of spins in the dual classical clock model $\{z_i\}$. The latter is the sum of the total number $\sf F$ of $\{B_p\}$ operators (faces) and the number $\sf V$ of $\{A_s\}$ operators (vertices). An additional degeneracy factor of $q^2$ in $N_{{\bf g},{\sf TC}}$ originates from the two constraints of (\ref{SPPMConstrd}) rendering two additional $\mathbb{Z}_q$ operators redundant. 

{\it Excitations of the {\rm SPPM}.} 
The duality (\ref{map1}) allows us to find {\it all} SPPM excitations through their duals in the classical clock model. 
As we noted earlier, for \(J,g >0\), any excited state (having stars for which ${\sf Eigenvalue}[A_s]\neq1$ and/or plaquettes ${\sf Eigenvalue}[B_p]\neq 1$) is also an excitation of the  ${\mathbb{Z}}_{q}$ TC model of Eq. (\ref{ZqTC}). 
Notably, different from other systems exhibiting  TC ground-states \cite{Gapless}, the identical spectra of the  SPPM and classical clock model are trivially quantized (given by sums of integer multiples of the coupling constants times cosines of the $q$ allowed discrete clock angles). Thus,  in particular, the  SPPM is gapped allowing for stable anyon excitations. 
An anyon in the \(\mathbb{Z}_q\) TC model, 
\( \epsilon^{{\sf n},{\sf \ell}}\)  
with $0 \le {\sf n}, {\sf \ell} \le (q-1)$, 
is a composite of ${\sf n}$ minimal ``electric'' charges ${\rm e}$ on the lattice sites 
(here, \( {\rm e} = \epsilon^{1,0}\)) and $\ell$ basic ``magnetic'' charges ${\rm m}$ (with \({\rm m} = \epsilon^{0,1}\)) associated with dual-sites. 
The energy of different TC eigenstates  depends only on anyon type and number. 
By contrast, in the SPPM, even the energies of the lowest-energy excitations may depend on the geometry, e.g., whether the anyons are nearest-neighbors.  

While, from our duality, the anyons \( \epsilon^{{\sf n},{\sf \ell}}\) of the TC models (with their properties well described by a Chern-Simons type TQFT \cite{Simon}) are also excitations of the SPPM (any excitation of the SPPM is a spatial composite of these), the description of the lowest-energy SPPM states may  generally go beyond that of these individual fundamental charges. 
Indeed, in general applications of the duality of Eq. (\ref{map1}), 
{\it lowest-energy} (gapped) excited states of the classical spin system may involve {\it non-compact} configurations such as the lowest-lying excited states of spin-glass \cite{SG1,ZED}, ANNNI \cite{ANNNI}, and other models. When the lowest-energy excitations of these classical systems are spatially non-compact, the duality (\ref{map1}) implies that the corresponding lowest-energy excitations of the quantum system are, 
rich composites of many point charges. A simple example is the SPPM  (\ref{SPPMHamiltonian}) on a torus with $J_{sp}=J>0$ and vanishing  fields $g_s=g_p=0$ at all sites $s$ and plaquettes $p$ except for $s=0$, where $8
J>g_{s=0}>0$. Here, the classical system has the ferromagnetic ground-state $z_{i}=1$ corresponding to conventional TC TQO on the quantum side of the duality (\ref{map1}) where $ \forall s, p$: ${\sf Eigenvalue}[A_{s}] = {\sf Eigenvalue}[B_{p}]=1$ with, on the torus, a $N_{{\bf g},{\sf TC}}= q^{2}$-dimensional ground-state manifold satisfying Eq. (\ref{KitaevCond}). The lowest-energy excited eigenstates  
correspond to different uniform classical clock model ferromagnetic states (and, thus, uniform dual $A_s$ and $B_p$ eigenvalues) associated with a global rotation by $2 \pi/q$ \footnote{
 Relative to the ground-state, the energy penalty of global $(2 \pi/q)$ rotations stems from the difference between the resulting $A_{s=0}$ eigenvalue (associated with an ensuing phase factor of $(2 \pi/q)$) with that favored by the external longitudinal field $g_{s=0}$ (i.e., an $A_{s=0}$ eigenvalue of unity suffering no such rotation). When $g_{s=0} > 8J >0$, the latter penalty will exceed that of states having a $(2 \pi/q)$ phase factor 
 of only two eigenvalues of either the star or the plaquette operators (or of two classical clock spins lying on the same sublattice in the dual model) with all other $A_s$ and $B_p$ eigenvalues (respectively, classical clock spin variables $z_i$) being $(+1)$. The above factor of eight in $8J$ arises from the product of two (the minimal number of disjoint $s \neq 0$ star or plaquette operator eigenvalues (classical clock spin variables) that may differ from unity given constraint  (\ref{SPPMConstrd})) and four (the length of the smallest square lattice domain wall around each of these rotated $s \neq 0$ star or plaquette eigenvalues (respectively, classical clock spins)).}. on a square lattice of even linear size \footnote{This parity requirement is necessary in order to ensure that Eq. (\ref{SPPMConstrd}) is satisfied.} that is endowed with periodic boundary conditions, when $q=2$  the $N_{{\bf g},{\sf TC}}$ degenerate lowest-energy excited eigenstates of the SPPM are the highest-energy eigenstates of the TC ($\forall s,p:$ ~${\sf Eigenvalue}[A_{s}] = {\sf Eigenvalue}[B_{p}]=-1$). These eigenstates correspond to {\it maximally dense alternating ${\rm e}$ and ${\rm m}$ charges covering the entire lattice}.  
 For \(g_s=g_p=0\), this excited state is degenerate with the conventional TC ground states displaying 
 a \(\mathbb{Z}_q\) global symmetry 
 \cite{SM}. When 
 \(g_s,g_p\) are non-zero (even if infinitesimal), 
 this global symmetry is lifted with, as discussed above, the lowest-energy excited state having maximally dense anyons.
 Since the anyons in this lowest excited state tessellate all of space, 
realizing pristine long-distance operations involving only individual anyons may be physically challenging. Rich relations exist between TQO symmetries, and resulting excitations, not limited to symmetry-enriched TC and unusual fracton models \cite{You_SymmFractonMatter,Wu_SymmEnrichToricCode_PhysRevB.108.115159,Wang_SymmEnrichToricCodeSolvable_PhysRevB.106.115104,kibe2024stabilizer}. We emphasize the duality as a description for the SPPM.

 {\it Conclusions.} 
In the current work we illustrated that various models displaying TQO \cite{SM} (satisfying the ground-state indistinguishability condition of Eq. (\ref{KitaevCond})) exhibit excitations whose description requires information beyond that provided by conventional TQFTs. To demonstrate this, we introduced 
theories 
containing distance dependent interactions (emulating perturbations) between anyons of known parent TC models. Such anyon interactions are expected in experimental realizations. 
We %
illustrated that 
particular TQO  \(\mathbb{Z}_q\) systems are dual to classical clock type models, with duals to other classical models easily constructed. Unlike thermally fragile \cite{Nussinov_2008_3DTC} TC and other topological models \cite{Weinstein_2019} that map onto one-dimensional systems, our models 
are identical to conventional high-dimensional Landau-type theories. 
Simple interactions augmenting those of well-studied topological systems such as the TC may change their spectra (while still leaving the systems topologically ordered). In particular, the lowest-energy excited states of the resulting topologically ordered systems {\it are not} necessarily those of single anyons for which conventional TQFT considerations apply nor bound states of such anyons nor screening on any finite length scale. Instead, in the thermodynamic limit, the {\it lowest-energy excitations} may correspond to an {\it infinite size dense lattice of anyons} that extends everywhere in space. 
There are illuminating connections between quantum error correcting (surface) codes and TQO as established by writing Hamiltonians in terms of stabilizer group generators \cite{Manny2001}, and encoding code words in ground-states. 
Our models include elements of the stabilizer group  (e.g., the products  $\{(A_sB_p)\}$), augmenting independent ($\{A_{s}\}, \{B_{p}\}$) generators. This inclusion leads to new physics. Indeed, as we  emphasized, the \(J=0\) SPMM (i.e., \(\mathbb{Z}_q\) TC model) is dual to decoupled one-dimensional chains  \cite{Nussinov_2008_3DTC,nussinov_ortiz_2009_IsingChains,PNAS-top} while for $J \neq 0$ (when products $\{(A_sB_p)\}$ appear in $H_{\sf SPPM}$), no such dimensional reduction results; the resulting dual high--dimensional classical systems may display large free energy barriers rendering them more immune to thermal fluctuations. 
In \cite{SM}, we expound on similar generalizations of string-net and other (including higher-dimensional) models 
their various features, and comparisons to certain theories \cite{comp1,comp2,comp3,cordova2022neutrino,cordova2022symmbreaking}. Our models realize quantum codes not explored to date.


 {\bf Acknowledgments.}
We are indebted to conversations with  Erez Berg, Jean-Noel Fuchs, Alexander Seidel, Steve Simon, Ruben Verresen, and Julien Vidal. Z.N. is grateful to the Leverhulme-Peierls senior researcher Professorship at Oxford supported by a Leverhulme Trust International Professorship grant [number LIP-2020-014]. This research was undertaken thanks in part to funding from the Canada First Research Excellence Fund. G.O. gratefully acknowledges support from the Institute for Advanced Study. This work was further performed in part at the Aspen Center for Physics, which is supported by National Science Foundation grant PHY-2210452.

\bibliography{citations}

\newpage

\newpage

\onecolumngrid

\section*{Supplemental Material:\\
Topological Orders 
Beyond Topological Quantum Field Theories}
\hspace*{5.5cm} P. Vojta, G. Ortiz, and Zohar Nussinov$^*$
\vspace*{0.5cm}

\twocolumngrid
In what follows, we present additional analyses that augment the more central considerations and findings presented in the main paper.


\section{Further Details of Duality Mappings of the SPPM to Classical Clock Models}

\subsection{Mapping an SPPM on an Open Surface to a Classical Clock Model}

We begin by considering the duality mapping between an ``open boundary condition'' planar SPPM to a square lattice classical clock model. In this mapping, clock spins of the classical model are duals of the star or plaquette terms of the SPPM. Within the system bulk, the SPPM star and plaquette operators are given by the four spin products  of the main text. On the boundary of the planar SPPM system, we define star and plaquette operators to include only those links that still lie in the lattice 
(and thus may include fewer than four spins). The products \(\prod_sA_s\) and \(\prod_pB_p\) will now act on boundaries and are unconstrained (do not satisfy Eq. (\ref{SPPMConstrd})). The result is a dual classical clock model on the same lattice with a number of clock model spins that is equal to that of the number of bonds in the quantum system (Eq. (\ref{map1})). If the number of independent bonds (generators of the bond-algebra),  i.e., the sum of the number of star and plaquette terms, $(N_s+N_p)$ is smaller than the number of links, there will be an additional degeneracy factor of $q^{(N_{\Lambda}-N_s-N_p)}$. For simple boundaries, this will result in a global ``holographic degeneracy'' \cite{Vaezi_2016}- a degeneracy scaling exponentially in the system boundary length. 

\subsection{Mapping an SPPM on a Torus to a Classical Clock Model}
 
\begin{figure}[htb]
\includegraphics[height=35ex]{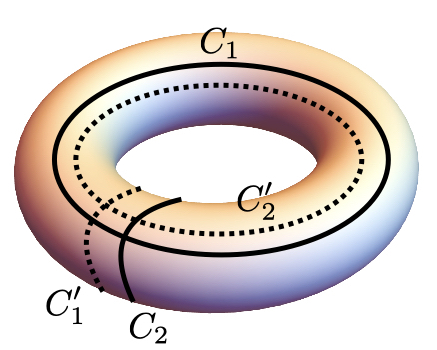}
\caption{The $d=1$ topologically distinct loops of the $\mathbb{Z}_q$ TC (associated with the symmetry operators of Eq. (\ref{SymmetryLoops})). The $Z^{\sf q}_{1(2)}$ symmetries (that constitute  logical operators of the TC) are given by products of Pauli matrices on the solid lines $C_{1,2}$ that pass through the lattice sites. The dashed lines $C_{1,2}^{\prime}$ inhabit the dual lattice and host the logical operators $X^{\sf q}_{1(2)}$.}
\label{1DSymmetries}
\end{figure}

 We now examine the duality between the SPPM on a torus (periodic boundary conditions) and a classical clock model. The incorporation of the periodic boundary condition constraint of Eq.(\ref{SPPMConstrd}) 
 can be achieved by
 amending the classical clock model Hamiltonian $H_{\sf c}$ of Eq. (\ref{HCLOCK}) by a finite number of bounded non-local terms (that sum to  $\delta H_{\sf c}$). 
 Towards this end, for any particular even ($\overline{s}$) and odd ($\overline{p}$) sublattice sites, we define the two operators
\begin{eqnarray}\label{nonlocals}
 \mathcal{W}_{\tilde{s}} \equiv \prod_{j\; {\sf even}, j\neq \tilde{s}} \!\!\! z_j, ~ \ \
\mathcal{W}^*_{\tilde{p}} \equiv \prod_{k\; {\sf odd}, k\neq \tilde{p}} \!\!\! z^*_k.
\end{eqnarray}
 We next implement the mapping of Eq. (\ref{map1}) with the modification that for the sites $\tilde{s}$ and $\tilde{p}$, 
\begin{eqnarray}
\label{map2}
A_{\tilde{s}} \leftrightarrow \!\!\! \prod_{j\; {\sf even}, j\neq \tilde{s}} \!\!\! z_j = \mathcal{W}_{\tilde{s}}, ~\ \
B_{\tilde{p}} \leftrightarrow \!\!\! \prod_{k\; {\sf odd}, k\neq \tilde{p}} \!\!\! z^*_k = \mathcal{W}^*_{\tilde{p}}.
\end{eqnarray}
With the mapping of Eq. (\ref{map2}), the product of all clock spin variables in the respective even and odd partitions will satisfy the global SPPM constraint of Eq. (\ref{SPPMConstrd}). Thus, out of the $N_\Lambda$ classical clock model spins, one pair of spins become redundant (fixed by the products of spins on other sites as on the righthand side of Eq. (\ref{map2})) and will not impact the energy of the $N_\Lambda$ spin model. That is, by comparison to the $N_\Lambda$ site classical clock model, there is an additional uniform global degeneracy factor of $q^2$ of each level (i.e., a factor of $q$ for each of the two classical spins at sites $\tilde{s}$ and $\tilde{p}$ that will no longer impact the system energy once the substitution of Eq. (\ref{map2}) is implemented into the system Hamiltonian). This degeneracy can also be read off from the two \(\mathbb{Z}_q\) symmetries of the system (Eq. (\ref{SymmetryLoops})) highlighted in Fig. \ref{1DSymmetries}. A detailed list of operators in the two models connected by this particular duality is given in Table \ref{MapTable}. 

We wish to find operators having an isomorphic action on the respective bond-algebra. Towards this end, consider an on-site clock rotation at site \(k\) in the clock model, 

\begin{eqnarray}
\label{x:eq}
x_k: z_j \leftrightarrow \begin{cases} e^{i2\pi/q}z_j & \mbox{if } j = k \\ z_j & \mbox{otherwise} \end{cases}.
\end{eqnarray}

If \(z_j \leftrightarrow A_s\) under the inverse duality map, then the respective transformation for the star operators reads
\begin{eqnarray}
 A_l \leftrightarrow \begin{cases} e^{i2\pi/q}A_l & \mbox{if } j = k \\ A_l & \mbox{otherwise} \end{cases}.
\end{eqnarray}
The transformation for the plaquette operators is analogous. Certain operators with this action exist but depend heavily on boundary conditions. For the periodic SPPM discussed in the main text, with the modifications of Eq. (\ref{deltaHc}), 
we notice that \(x_k\) also rotates one of the respective \(\mathcal{W}\)'s. That is, the \(x_k\) operators act like (and are in fact dual to) \(Z^{\sf q}_{\tilde{s},s}\) and \(X^{\sf q}_{\tilde{p},p}\).
\begin{eqnarray}
    Z^{\sf q}_{\tilde{s},s} = \prod_{j\in C_{\bar{s},s}}Z_j, \quad  
    X^{\sf q}_{\tilde{p},p} = \prod_{j\in C'_{\bar{p},p}}X_j.
\end{eqnarray}
Here, \(C_{\tilde{s},s}\) is a contour in the lattice with endpoints at \(s\) and \(\tilde{s}\) while \(C'_{\tilde{p},p}\) is similarly placed in the dual-lattice with endpoints at \(p\) and \(\tilde{p}\).

Given a conventional classical clock model subject to   periodic boundary conditions, consider a duality to an SPPM with the modifications of  Eq. (\ref{OmegaAB}). 
We next construct the dual string products 
\begin{eqnarray}
\label{z:bar}
    \bar{Z}^{\sf q}_{\bar{s},s} = x^{-1}_{\Omega_1}\prod_{j\in C_{\bar{s},s}}Z_j, \quad  
    \bar{X}^{\sf q}_{\bar{p},p} = x^{-1}_{\Omega_2}\prod_{j\in C'_{\bar{p},p}}X_j.
\end{eqnarray}
Here, we amended the string products with counter-rotating operators on \(\Omega\) so as to leave the operators \(\bar{A_s}\) and \(\bar{B_p}\) of Eq. (\ref{OmegaAB})  unchanged,
For an SPPM on the infinite planar lattice or embedded on a manifold with a boundary, \(x_k\) is dual to a product along a string-like contour with this contour either diverging to infinity or terminating on the boundary, respectively. 

The \(J,g >0\) SPPM has a ground-state subspace identical to that of the TC ground-state subspace, \(V_{TC}\).  We know the classical clock model at \(g=0\) has a global symmetry \(\mathcal{U}_{clock}\). Removing \(g\)-terms from the SPPM also introduces a global symmetry into the system. The \(J>0,~g=0\) SPPM  ground-states are specified by the condition \(A_s\ket{\psi} = B^\dagger_p\ket{\psi}=e^{i2n\pi/q}\ket{\psi}\), \(n = 0,1,2,...,q-1\).  The respective \(n=0\) subspace is then that of the TC model, \(V_{TC}\). The global symmetry group generated by an operator 
\(\mathcal{U}_{sp}\), detailed below, then connects the \(q\) subspaces which correspond respectively to the \(q\) ground-states of the dual \(g=0\) clock model.

Specifically, the global rotational symmetry of a conventional clock model is
\begin{eqnarray}
  \mathcal{U}_{clock} = \prod_jx_j.
\end{eqnarray}

The dual symmetry is then generally the appropriate aforementioned product of strings, forming the global symmetry of the SPPM. For the periodic SPPM discussed in the main text (with the modifications of Eq. \ref{map2} in the dual clock model at \(\tilde{p}\) and \(\tilde{s}\)) the global symmetry is simply

\begin{eqnarray}
\label{usp}
   \mathcal{U}_{sp} = \Big(\prod_pX^{\sf q}_{\tilde{p},p} \Big)\Big(\prod_s Z^{\sf q}_{\tilde{s},s} \Big).
\end{eqnarray}
An alternative view of topological order centers on long-range entanglement \cite{Wen_2013_LightFermion}. Here, \(\mathcal{U}_{sp}\) is a tensor product, leaving the known long-range entanglement features of \(V_{TC}\) unchanged. The two factors of \(\mathcal{U}_{sp}\) also have natural interpretations in the \(q=2\) (i.e., Ising) model. They connect the ferromagnetic ground-states to the ground-states of its antiferromagnetic phase. The global \(X\)-string product in Eq. (\ref{usp}), which generates \(m\)-type anyons on all plaquettes, is dual to an operator which globally rotates all classical clock model spins on sites of the odd sublattice. 

   Performing the duality of Eqs. (\ref{map1}, \ref{map2}) to the SPPM Hamiltonian of Eq. (\ref{SPPMHamiltonian}) 
   with periodic boundary conditions (where Eq. (\ref{SPPMConstrd}) applies) leads to a classical spin model with the said Hamiltonian of $(H_{\sf c} + \delta H_{\sf c})$. The precise form of $\delta H_{\sf c}$ depends on whether \(\tilde{s}\) and \(\tilde{p}\) are nearest-neighbors. When, e.g., $\tilde{s}$ and $\tilde{p}$ are well separated, 
\begin{eqnarray}
\label{deltaHc}\nonumber 
 \delta H_{\sf c} = -J\sum_{\langle j ~\tilde{p}\rangle} {\mathcal{W}^*_{\tilde{p}}z_{j}+\mathcal{W}_{\tilde{p}}z^*_{j}}-g(\mathcal{W}_{\tilde{p}}+\mathcal{W}^*_{\tilde{p}}) \nonumber
\\-J\sum_{\langle k~\tilde{s}\rangle}{\mathcal{W}^*_{\tilde{s}}z_{k}+\mathcal{W}_{\tilde{s}}z^*_{j}} -g(\mathcal{W}_{\tilde{s}}+\mathcal{W}^*_{\tilde{s}}) \nonumber
\\ +J\sum_{\langle i~\tilde{p}\rangle} {z^*_{\tilde{p}}z_{j}+z_{\tilde{p}}z^*_{j}}+g(z_{\tilde{p}}+z^*_{\tilde{p}}) \nonumber
\\+J\sum_{\langle k~\tilde{s}\rangle}{z^*_{\tilde{s}}z_{k}+z_{\tilde{s}}z^*_{j}} +g(z_{\tilde{s}}+z^*_{\tilde{s}}).
 \end{eqnarray}
 This Hamiltonian is a function of $(N-2)$ independent spins once Eq. (\ref{map2}) is implemented. 
Thus, on the 4-fold coordinated square lattice, \(| \delta H_{\sf c} | \le (32J + 8g)\) (here the prefactor of $32$ originates from the product of $8$ (number of relevant terms) $\times~ 4$~ (lattice coordination number)). An individual clock spin (at an arbitrary site $j$) can be written as \(z_j = e^{2i\pi k_j/q}, k_j = 0,1,...,q-1\). 
With this substitution, the clock model Hamiltonian of Eq. (\ref{HCLOCK}) then  reads 
\begin{eqnarray}
H_{\sf c} = -2J\sum_{\langle i,j\rangle}\cos(\frac{2\pi(k_i-k_j)}{q})-2g\sum_j\cos(\frac{2\pi k_j}{q}).
\end{eqnarray}
In what follows, we first briefly discuss the $J=0$ case (the \(\mathbb{Z}_q\) TC model) and then turn to the SPPM ($J\ne 0$).

For \(J=0\), we may employ a trivial generalization of the duality mapping of \cite{nussinov_ortiz_2009_IsingChains} so that the \(\mathbb{Z}_q\) TC model (\(J=0\)) is dual to a pair of uncoupled periodic \(\mathbb{Z}_q\) clock model  spin chains. In this $J=0$ limit, denoting the clock spin variables in the first periodic chain by \(z^{(1)}\)  and those in the second chain by \(z^{(2)}\) gives the dual classical clock model Hamiltonian
\begin{eqnarray}
    H_{\sf c}^{\sf chains} = -g\sum_{i=1}^N (z^{(1)*}_iz^{(1)}_{i+1}+z^{(1)}_jz^{(1)*}_{j+1})\nonumber
    \\-g\sum_{j=1}^N (z^{(2)*}_jz^{(2)}_{j+1}+z^{(2)}_jz^{(2)*}_{j+1}).
\end{eqnarray}
Here, the corresponding partition functions are quite simple to calculate via the transfer matrices of the model. The discrete clock rotation invariant transfer matrix may be diagonalized by a discrete Fourier transform. For the single chain, the partition function reads
\begin{eqnarray}
\label{Fourier}
\hspace*{-0.5cm}
\mathcal{Z}^{\sf chain}_{\sf c}= \sum_{\tilde n=1}^q\Big (\sum_{n=1}^q \exp{(2g\cos(\frac{2\pi n}{q})+i\frac{2\pi n\tilde n}{q})}\Big)^N.
\end{eqnarray}
In the \(N \to \infty\) limit, the non-oscillatory \(\tilde n=0\) term dominates over all other Fourier components.   in Eq. (\ref{Fourier}),   
\begin{eqnarray}
\label{ZcN}
    \mathcal{Z}^{\sf chain}_{\sf c} \approx \Big ( \sum_{n=1}^q(\exp{(2g\cos(\frac{2\pi n}{q}))}\Big)^N.
\end{eqnarray}
 Thus in the thermodynamic limit, the partition function of the \(J=0\) SPPM model (\(\mathbb{Z}_q\) TC model) which is the product of partition functions of two identical chains, \(\mathcal{Z}^{\sf chain}_{{\sf c},1} \times \mathcal{Z}^{\sf chain}_{{\sf c},2}\), is given by the square of Eq. (\ref{ZcN}).
 
We now return to the general case of arbitrary $J$ (the SPPM). Here, the duality implies that the free energy density of the SPPM whose classical dual is given by $(H_{\sf c} + \delta H_{\sf c})$ is identical to that of the conventional clock model of $H_{\sf c}$ . As we discussed above, taking into account the constraints of Eq. (\ref{map2}), there is a global degeneracy of $q^2$ of each level.  The \(q \to \infty\) limit yields a duality to a classical XY model.

\begin{table}[h!]
\begin{center}
\begin{tabular}{ |c|c|c| }
 \hline
 SPPM & Clock Model \\ 
  \hline
 \(A_s\) & \(z_i\) (even) \\ 
 \(B_p\) & \(z^*_j\) (odd) \\
 \(\prod_sA_s\) & \(\mathbb{1}\)\\
 \(\prod_pB_p\) & \(\mathbb{1}\)\\
 \(X^{\sf q}_{1(2)}\) or \(Z^{\sf q}_{1(2)}\) & \(\mathbb{1}\)\\
 \(Z^{\sf q}_{s_1,s_2}\) & \(x_{s_1}x_{s_2}\)\\
 \(X^{\sf q}_{p_1,p_2}\) & \(x_{p_1}x_{p_2}\)\\
 \(A_{\tilde{s}}\) & \(\prod_{j\; even, j\neq \tilde{s}}z_j\) \\
 \(B_{\tilde{p}}\) & \(\prod_{k\; odd, k\neq \tilde{p}}z^*_k\)\\
\(X^{\sf q}_{\tilde{p},s(x_i)}\) or \(Z^{\sf q}_{\tilde{s},p(x_i)}\) &  \(x_i\)\\
 \(\mathcal{U}_{sp}\) & \(\prod_ix_i\)\\
 \hline
\end{tabular}
\end{center}
\caption{Correspondences between the operators in the  SPPM of Eqs. (\ref{SPPMHamiltonian}, \ref{SymmetryLoops}, \ref{z:bar}) and clock model operators of Eqs. (\ref{HCLOCK}, \ref{x:eq}).}
\label{MapTable}
\end{table}
\section{Mapping Classical Clock Models to SPPM systems}

We now discuss the ``inverse'' of the above mapping and illustrate how to map a conventional classical clock model on a torus onto a modified SPPM type system. For periodic boundary conditions, the SPPM  satisfies the constraint of Eq. (\ref{SPPMConstrd}). 
Previously, we discussed the mapping from the SPPM to the clock model while preserving this constraint. Now we examine a map from the classical spin model to the SPPM which preserves looser constraints. We will see that each link-local operation \(X\) or \(Z\) corresponds to flipping a pair of diagonally neighboring clock spins. This enforces the dual constraint of pairwise flipping spins and preserving the spin parity number. However this constraint is not valid in the clock model, where a single clock spin can always be rotated. So conventional clock models with periodic boundary conditions will not map into SPPMs with periodic  boundary conditions unless we include additional degrees of freedom. We next consider various types of periodic clock models and their SPPM duals.

\subsection{``Even-Sized'' Square Lattices}

We first consider a classical clock model on a  square lattice (of size $L_1 \times L_2$) on a torus and map it into an SPPM-like model on a lattice that is endowed with periodic boundary conditions. When both $L_1$ and $L_2$ are even,  the classical spin model vertex to SPPM star/plaquette operator mapping (an inverse of Eq. (\ref{map1})) may be applied everywhere relating the original square lattice on which the classical clock model is defined to the $45^{\circ}$ rotated square lattice on which an SPPM type model is defined (see Fig. \ref{LatticeMapFig}). This mapping will result in \(\prod_sA_s\) and \(\prod_pB_p\) being unconstrained, unlike Eq. \ref{SPPMConstrd}. To 
make an analog of these products unconstrained in the SPPM like model, we introduce two independent \(\mathbb{Z}_q\) operators \(\Omega_{1}\) and \(\Omega_{2}\). With these, we replace the individual star and plaquette operators $A_s$ and $B_p$ in Eq. (\ref{SPPMHamiltonian}) by, respectively, 
\begin{eqnarray}
\label{OmegaAB}
\bar{A_s} \equiv \Omega_1A_s ,~ \bar{B_p} \equiv \Omega_{2}B_p.
\end{eqnarray}

The operators \(\Omega_{1}\) and \(\Omega_{2}\) correspond to two fully decoupled \(\mathbb{Z}_q\) redundant gauge-like degrees of freedom leading to a 
 \(q^2\)-fold degeneracy.

\subsection{``Odd-Sized'' Square Lattices}
\begin{figure*}
\subfloat[The operator \(\mathcal{B}_0\) at the intersection of two defects.]{\includegraphics[scale = 0.2]{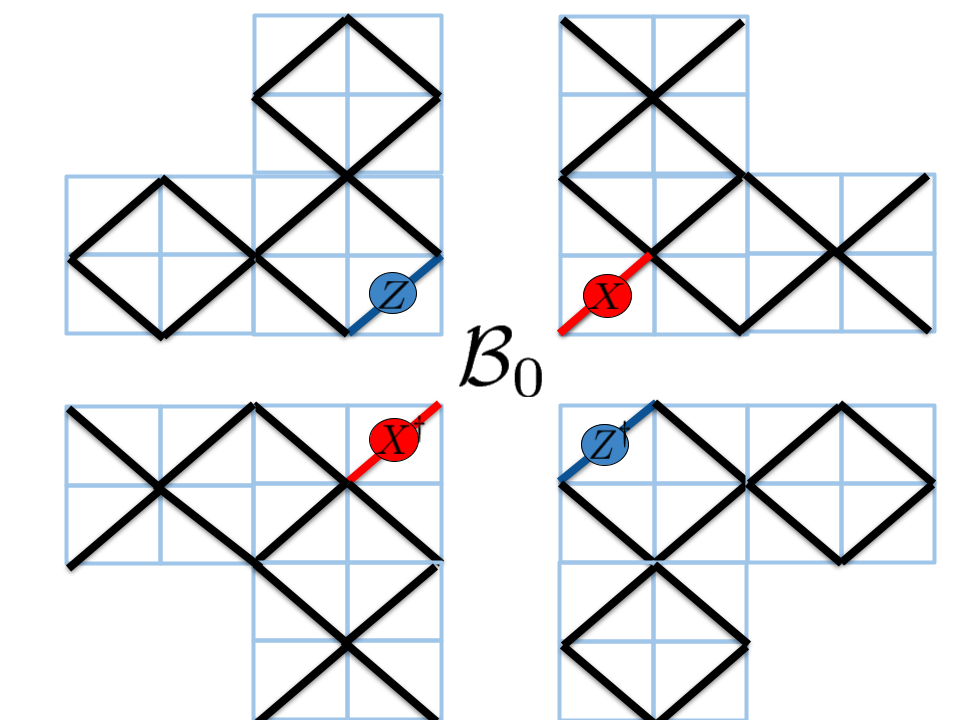}}
\subfloat[The operators \(\mathcal{B}_+,\mathcal{B}_-\) along a single defect,]{\includegraphics[scale = 0.2]{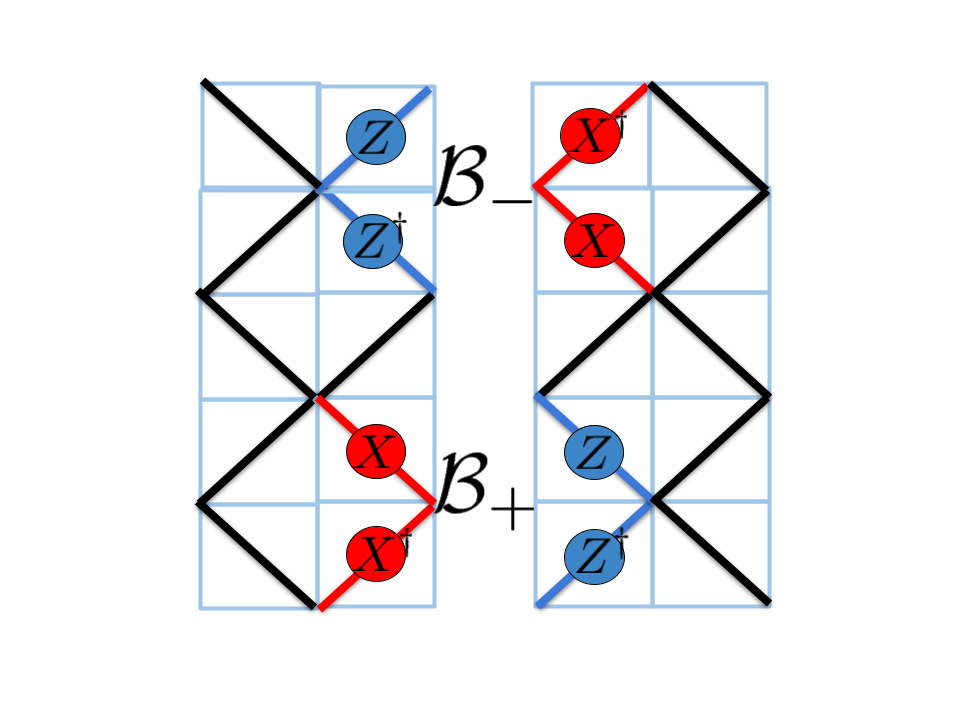}}
\caption{(Color online.) Stitching operators \(\mathcal{B}_0, \mathcal{B}_+, \mathcal{B}_-\) obey the same general bond-algebraic constraints. They can be used to pass strings/anyons across the defects. The SPPM lattice is in black and the corresponding \(\mathbb{Z}_2\) clock lattice is in light blue.}
\label{NonBipartiteOperatorFIG}
\end{figure*}

We now turn to duality mappings from a classical clock model on a torus to an SPPM type system for square lattices when, at least, one of the sides $L_1$ or $L_2$ is odd.  Here, the constructs  of 
Eq. (\ref{OmegaAB}) may still be applied, allowing for the desired lack of constraints \ref{SPPMConstrc}. 
In this case, the geometrical implementation of Eq. (\ref{map2}) (Fig. (\ref{LatticeMapFig})) will not be consistent at the boundaries of the square lattice on which the classical clock model is defined. We can resolve this by 
constructing boundary operators \(\mathcal{B}_i\)  
that will obey constraints equivalent to those in Eq. \ref{SPPMConstra}-\ref{SPPMConstrd}. Towards that end, we introduce new operator products \(\mathcal{B}_{+/-}\) that act on both sides of the ``boundary" as shown in Fig.  \ref{NonBipartiteOperatorFIG}. We further introduce a corner operator \(\mathcal{B_0}\) if both \(L_1\) and \(L_2\) are odd. 
These boundary terms are now further included in the amended  Hamiltonian of Eq. (\ref{SPPMHamiltonian}) with the replacement of star and plaquette terms following Eq. (\ref{OmegaAB}). Neighboring \(\mathcal{B}_i\) overlap by one link and all \(\mathcal{B}_i\) map individually into the clocks to which the disconnect has been associated. These disconnects generally replace the constraint of Eq. (\ref{SPPMConstrd}) by the looser condition of \(\prod_{s}A_s\prod_{p}B_p\prod_{\forall \mathcal{B}}\mathcal{B} = 1\). However, 
with modifications of Eq. \ref{OmegaAB}, the respective global products are unconstrained.

 \subsection{Symmetries and the Toric Code Anyon Algebra}
In the TC,  an application of \(\mathcal{U}_{sp}\) on the ground-state  generates anyons uniformly at every site and dual-site (i.e., plaquette center) of the lattice. The SPPM nearest neighbor coupling (\(J\) type) terms do not distinguish between the TC vacuum \(\ket{k,0}\) and anyon populated states \(\ket{k,n} = \mathcal{U}^n\ket{k,0}\) where  \(k = 1,2,...,q^{2({\sf genus})}\) and \( n = 0,1,...,q-1\). 
For these states, 
\(A_s\ket{k,n} = e^{i2n\pi/q}\ket{k,n}\) and \(B_p\ket{k,n} = e^{-i2n\pi/q}\ket{k,n}\). 
With the $\epsilon^{\sf n,\ell}$ excitation taxonomy of the main text, the action of 
\(\mathcal{U}_{sp}\), 
\begin{eqnarray}
\mathcal{U}_{sp}:\;\; \epsilon^{\sf{n,\ell}} \mapsto \epsilon^{\sf{n +1,\ell+1}}.
\end{eqnarray}
Thus, 
\begin{eqnarray}
 \mathcal{U}_{sp}(\epsilon^{\sf{n,\ell}} \times \epsilon^{\sf{u,v}}) = \mathcal{U}_{sp}(\epsilon^{\sf{n +u,\ell+v}}) = \epsilon^{\sf{n+u+1,\ell+v+1}},   
\end{eqnarray}
while 
\begin{eqnarray}
  \mathcal{U}_{sp}(\epsilon^{\sf{n,\ell}} ) \times \mathcal{U}_{sp}(\epsilon^{\sf{ u,v}}) = \epsilon^{\sf{n+u+2,\ell+v+2}} \nonumber
  \\ \neq \mathcal{U}_{sp}(\epsilon^{\sf{n,\ell}} \times \epsilon^{\sf{u,v}}).  
\end{eqnarray} 
In other words, the TC excitation algebra 
does not remain invariant under a transformation by  \(\mathcal{U}_{sp}\). When \(g=0\), both the (i) vacuum and (ii) the maximally possible dense anyon state are ground-states of the SPPM. New, but related, descriptions are required to accurately capture the behavior of the excitations which in this model are merely the domain walls between differing regions locally forming a ground-state.

\section{SPPM Variants and Their Excitations}










\subsection{An Ashkin-Teller Model Dual} 

Another illustrative model in the breadth of SPPM variants is in considering the dual model to a $\mathbb{Z}_q$ Ashkin-Teller Model \cite{aoun2023phase_AshkinTeller}. The best known ($q=2$) variant of this model may be understood as an interacting pair of Ising models on (arbitrary) identical lattices. The model is given by Hamiltonian
\begin{eqnarray}
    H_{AT} = -J_1\sum_{\langle i,j \rangle} s_is_j-J_2\sum_{\langle i,j \rangle} s'_is'_j-U\sum_{\langle i,j \rangle} s_is_js'_is'_j.
\end{eqnarray}
The Ashkin-Teller model is known to display a line of fixed points \cite{PhysRevLett.26.832_BaxterLine}. Following our recipe of Eq. (\ref{map1}), each Ising spin may be mapped to a star or plaquette operator of a one of two distinct TC models. That is, all of the spins $\{s_i\}_{i=1}^{N}$ will be mapped to the TC plaquette and star operators associated with one copy of the lattice (a single ``layer'')  containing its $N$ lattice links while the spins $\{s'_{i}\}_{i=1}^{N}$ will be mapped onto the TC star and plaquette operators defined on another replicated copy of the same  lattice (or layer). On this two-layer lattice, 
the duality of Eq. (\ref{map1}) yields the Hamiltonian
\begin{eqnarray}
\label{ATD}
     H_{\sf{AT~dual}} = -J_1\sum_{\langle i,j \rangle} A_sB_p-J_2\sum_{\langle i,j \rangle} A'_sB'_p \nonumber
     \\ -U\sum_{\langle i,j \rangle} A_sB_pA'_sB'_p.
\end{eqnarray}
For positive coupling constants $J_{1,2}, U >0$, the ground-states
of $H_{{\sf{AT~dual}}}$ are those with $A_s=B_p=A'_s=B'_p=1$- i.e., the direct product of the TC ground-states on the two different copies of the lattice (or layers). By virtue of being the direct products of topological states that each satisfy Eq. (\ref{KitaevCond}), it follows within the ground-state space of $H_{\sf{AT~dual}}$, Eq. (\ref{KitaevCond}) will also be satisfied for any (quasi-)local operator on the two-layer system. 
Similarly, the considerations of  \cite{nussinov_ortiz_2009_IsingChains,PNAS-top,Nussinov_2008_3DTC} ensure that this model displays topological order also at higher energies. By our duality, the aforementioned fixed point line of the Ashkin-Teller appears unchanged for our dual model of Eq. (\ref{ATD}). There is thus a {\it continuum of distinct fixed points} that are all associated with topological order and host TC type anyons  (i.e., basic ``electric'' and ``magnetic'' anyons and their composites on the two layer lattice).
 
\subsection{Long-range models} 
Our construct is not limited to short-range interactions. Consider a long-range Ising model \cite{Hiley_Joyce_1965} governed by the Hamiltonian ($J>0$),  
\begin{equation}
\label{Isinglong}
H_{\sf LRI} = -\sum_{i<j}\frac{J}{|r_i-r_j|^{\zeta}}\sigma_{i}\sigma_{j}-g_i\sum_{i}\sigma_{i},
\end{equation}
with \(\sigma_i = \pm 1\).
In order for a ferromagnetic ground-state in $D$ dimensions to have a well defined finite energy density in the thermodynamic limit, the exponent $\zeta > D$.  
Repeating our earlier steps (in particular, those relating to the mapping of Eq. (\ref{map1})) 
yields a quantum dual given by 
\begin{eqnarray}\label{LongRangeSPPM}\nonumber
H_{\sf LRSP} 
= -\sum_{s,p}\frac{J}{|r_s-r_{p}|^{\zeta}}A_sB_p-\sum_{s\neq \tilde{s}}\frac{J}{|r_s-r_{\tilde{s}}|^{\zeta}} A_sA_{\tilde{s}}\\\nonumber
-\sum_{p\neq \tilde{p}}\frac{J}{|r_p-r_{\tilde{p}}|^{\zeta}}B_pB_{\tilde{p}}-g_s\sum_s{A}_{s}-g_p\sum_{p}{B_p}. \\
\end{eqnarray}
Once again, when $g_s,g_p,J>0$, this quantum model shares a ground-state space with the TC.
Excited eigenstates of $H_{\sf LRSP}$ are in a one-to-one correspondence with those 
of the TC albeit with different energy penalities. 
Pairs of anyons in this model will experience string tension described by the long-range Ising model of $H_{\sf LRI}$.

\subsection{Frustrated Topological Order}
In the main text, we emphasized the \(J>0, g>0\) form the SPPM, in which the ground-state space corresponds to ferromagnetic ground-state in the classical model. When \(J<0\) and \(g>0\), the classical model exhibits an antiferromagnetic ground-state when \(J<-g/4\). The line \(J=-g/4\) then describes the boundary between ferromagnetic and antiferromagnetic phases. In the main text we illustrated that each classical spin configuration corresponds to a \(N_{{\bf g},{\sf TC}}\)-dimensional sector of the SPPM. In this way, the N\'eel state dual sectors together form the space

\begin{eqnarray}
 \mathcal{H}_{\sf Neel}= \Bigl\{ \ket{\psi} ; A_sB_p\ket{\psi}=-\ket{\psi}\Bigl\}.
\end{eqnarray}

This sector can be constructed by operating on ferromagnetic spaces \({\cal H}_{{\sf Ferro+}}\) or \({\cal{H}}_{\sf{Ferro-}}\) with one of the following operators.

\begin{eqnarray}
   X_{\sf Neel} = \prod_pX^{\sf q}_{\tilde{p},p} \\
   Z_{\sf Neel} = \prod_s Z^{\sf q}_{\tilde{s},s}.
\end{eqnarray}
The operator \(\mathcal{U}_{sp}\) of Eq. \ref{usp} then connects N\'eel state dual sectors while preserving topological winding numbers associated to the symmetries of Eq. \ref{SymmetryLoops} and Fig. \ref{1DSymmetries}. The product of \(X_N\) and \(Z_N\) is \(\mathcal{U}_{sp}\). Two subspaces of \(V_N\) can be identified by the value of \(A_0\), (in principle an arbitrary star but we take it at the origin as in the previous subsection) which we write as

\begin{eqnarray}
   \mathcal{H}_{\sf Neel-}= \Bigl\{ \ket{\psi} ; A_sB_p\ket{\psi}=A_0\ket{\psi}=-\ket{\psi},\Bigl\}\\
   \mathcal{H}_{\sf Neel+}=\Bigl\{\ket{\psi} ; A_sB_p\ket{\psi}=-A_0\ket{\psi}=-\ket{\psi}\Bigl\}.
\end{eqnarray}

Not all  \{\((J_{sp}A_sB_p)\}\) terms in the Hamiltonian can be simultaneously minimized (so as to yield a  ``frustration free'' system). This may not be possible for certain lattices as is easily demonstrated by the duality of 
Eq. (\ref{map1}) to a frustrated classical model to generate a quantum  plaquette/star model counterpart. We may interpret such a frustration as a modification on the typical TC model in which anyons occur in even pairs (or fused products with the total number of generating anyons still even). If we impose periodic boundary conditions on the classical clock model, frustration may occur along either identified boundary, depending on if \(L_1,L_2\) are even or odd. This may be resolved by twisting the boundary conditions and attaching a minus sign to \(J\) along the frustrated boundary. 

\subsection{Non-compact Excitations}

A simple example of a Hamiltonian in the bond-algebra of possible \(A_s,B_p\) couplings is discussed in the main text, in which the excitations are non-local colections of anyons. We provide a further example with strictly uniform couplings in the Hamiltonian (in \(\mathbb{Z}_2\) for simplicity),

\begin{eqnarray}
\label{HAABB}
    H_{\sf SSPPPM} = -K\sum_{links}A_{s1}A_{s2}B_{p1}B_{p2} \nonumber
    \\ -g\sum_s A_s-g\sum_p B_p.
\end{eqnarray}
For concreteness, in what follows, we will consider positive couplings, \(K,g \geq 0\). The first sum in the ``Star Star-Plaquette Plaquette Product Model'' 
of Eq. (\ref{HAABB}) is that over products of two star and two plaquette operators that are associated with each single link of the lattice. Such products map under our duality to 2$\times $2 spin plaquettes of Ising spins. Since the underlying bond-algebra generated by the operator set \(\set{A_s,B_p}\) is commutative, the underlying ground state space is still that of previously considered SPPM models (that of the TC model). For \(g=0\), global creations of \(e\)-type
or \(m\)-type anyons is a symmetry of the model (in the absence of any geometric frustration). For finite \(g\), the corresponding energy is \(2gN_\Lambda\). The energy of a local anyonic pair excitation is \(16K+2g\). It follows that when

\begin{eqnarray}
    N_\Lambda < 8\frac{K}{g}+2,
\end{eqnarray}
a global system spanning excitation is energetically favorable by comparison to that of the  thus far conventionally known excitations of local anyon excitations. That is the lowest excitation of the model are non-compact, global configurations of anyons. For a model with a boundary enabling single anyon creation, with excitation energy \(8K +g\), this condition becomes

    \begin{eqnarray}
    N_\Lambda < 4\frac{K}{g}+ 1.
\end{eqnarray}

\subsection{Star-Star and Plaquette-Plaquette Coupling}

The simplest case in which the lowest--energy excitations have altered braiding statistics is evident if we introduce the 
coupling to be between stars and plaquettes. When $q=2$ (the Ising case), the corresponding Hamiltonian reads 
\begin{eqnarray}
   H_{\sf ss,pp}=J\sum_{\langle s,s' \rangle }A_{s}A_{s'}-J\sum_{\langle p,p' \rangle }B_{p}B_{p'} \nonumber \\
   -g\sum_{s} A_{s}-g\sum_{p} B_{p}.
\end{eqnarray}

Note that nearest neighbor star (or  plaquette) operators share a common lattice link. Applying Eq. (\ref{map1}), it is seen that this Hamiltonian is dual to a pair of decoupled 2d Ising models. 
The lowest excitations of this model for \(J,g > 0\) are the bound electric \((ee)\) and magnetic \((mm)\) anyon pairs created by applying a single \(Z\) link operator (whence the two stars sharing that link are populated by $e$ charge) or single \(X\) link operator (creating $m$ charges on the two on plaquettes sharing the link) the ground- state respectively.  Splitting these pairs increases associated domain wall energy contributions by \(4J\). When \(q=2\), bound pairs braid trivially with one another, and have trivial winding numbers, unlike the free anyons.

\begin{figure*}
\subfloat[]{\includegraphics[height=35ex]{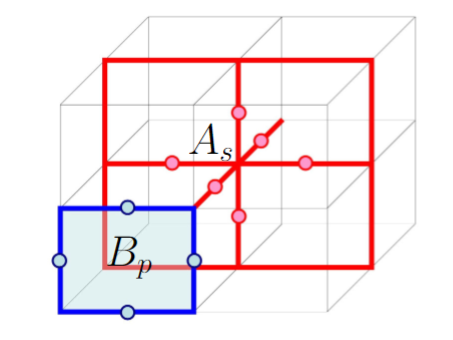}}
\subfloat[]{\includegraphics[height=30ex]{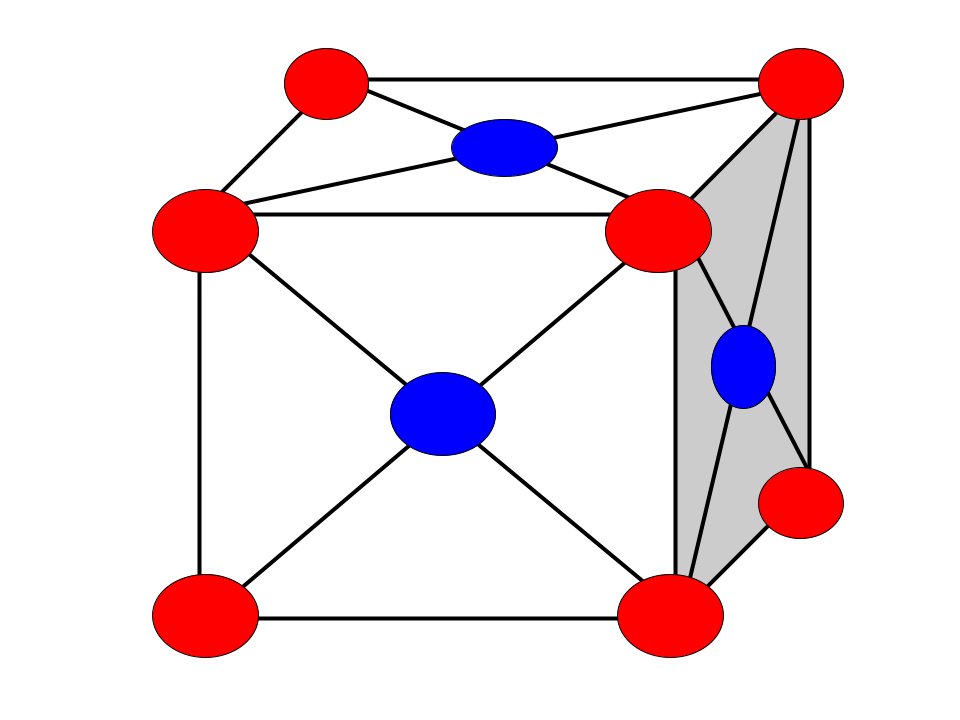}}
\caption{(Color online.) (a) Red circled links form a 3D toric code \cite{Nussinov_2008_3DTC} star while circled blue links form a 3D plaquette. (b) A cubic cell of the FCC. Plaquette operators map into blue vertex clocks on FCC faces. Star operators map to red vertex clocks on FCC corners.}
\label{3DStarPlaq}
\end{figure*}

\section{String-Net Models and Quantum Doubles}

Our illustration that a TQO system (the planar SPPM) is dual to a Landau type theory (the classical clock model) can be readily extended to many other models. We mention a few of these below. 
String-Net (SN) models use fusion category inputs to construct models on trivalent lattices \cite{Lin_2021_genStringNet}. Analogous to the TC model, the SN Hamiltonians may be expressed as sums of star and plaquette type operators,
  $ H_{\sf SN} = -\sum_{s}A^{\sf SN}_s-\sum_{p}B_p^{\sf SN}$.  
 The star operator \(A^{\sf SN}\) projects states of links sharing a common endpoint onto a space of allowed link configurations 
 while \(B_p^{\sf SN}\) transforms link configurations. 
  The definitions of \(B_p^{\sf SN}\) are given in terms of nontrivial \(F\)-symbols describing the transformations at each vertex \cite{Levin_2005}. We propose extensions of SN models that include \(A_s^{\sf SN}B_p^{\sf SN}\)  terms, as in Eq. (\ref{SPPMHamiltonian}), capturing nearest-neighbor anyon interactions. Again, this geometrical  dependence has no TQFT description and can be understood through dualities to classical systems. We treat Quantum Double models, which map to a subclass of SN models \cite{Buerschaper_2009}.
We specifically consider the \(\mathbb{Z}_q\) Quantum Double Hamiltonian \cite{Pachos_2012} ,
\begin{equation}\label{Qdouble_Hamiltonian}
    H_{\sf QD} = -\sum_{s}A^{\sf QD}_s-\sum_{p}B^{\sf QD}_p,
\end{equation}
(where \(A^{\sf QD}_s = \sum^{q-1}_{\sf r =0}A_s^{\sf r}\) and \(B^{\sf QD}_p=\sum^{q-1}_{\sf r =0}B_p^{\sf r}\) are sums through all powers of the \(\mathbb{Z}_q\) TC \(A_s\) and \(B_p\) operators) and its  extension 
\begin{equation}\label{SP_Qdouble_Hamiltonian}
    H_{\sf SPQD} = -J\sum_{\langle s~p \rangle} A^{\sf QD}_sB^{\sf QD}_p-g\sum_{s}A^{\sf QD}_s-g\sum_{p}B^{\sf QD}_p.
\end{equation}
Repeating, 
the mapping of Eq. (\ref{map1}) 
transforms the string-net operators to geometric series of the classical clock spin variables and vice versa, 
\(A^{\sf QD}_s  \leftrightarrow \sum
^{q-1}_{\sf r =0}z^{\sf r}_i\) and \(B^{\sf QD}_p \leftrightarrow \sum^{q-1}_{\sf r =0}
(z^{\sf r}_j)^*\),   
leading to an unconventional classical clock model that is dual to $H_{\sf SPQD}$. A possible star-plaquette coupling in quantum double models was also mentioned in \cite{KomarCardinal_PhysRevB.96.195150}, where the coupling introduces 
tunable anyon masses. 




For simplicity of notation, in what follows henceforth, we omit the superscript ``\({\sf SN}\).'' If the trivalent lattice three link labels connected to vertex \(s\) are in the state \(\ket{ijk}\), then \(A_s\ket{ijk} = \delta_{ijk}A_s\) with  \(\delta_{ijk} = 1\) for allowed vertex states and \(0\) for not allowed states. \(B_p\) is then a linear combination of transforms \(B_p^w\) 

\begin{equation}\label{StringNetPlaquette}
B_p = \sum_s \sum_w a_wB_p^w.
\end{equation}

The \(B_p^w\) are associated to a transformation at each vertex on a plaquette. A geometric intepretation is that \(B_p^w\) multiplies each plaquette with an  \(w\)-th oriented string loop. This view is sometimes called the ``fat lattice construction'' \cite{Simon}. The orientation dependence of these loops immediately gives the global constraint \(\prod_p B_p^w = 1, \; \forall w\). Consider a hexagonal plaquette state \(\ket{\psi^{ghijkl}_{abcdef}}\) (and legs) with leg links \(a,b,c,d,e,f\) and inner plaquette links \(g,h,i,j,k,l\) such that \(a\) forms a star with \(g\) and \( h\), \(b\) with \(h\) and \(i\),  etc., 

\begin{equation}\label{StringNetPlaqAction}
B_p^w\ket{\psi_p} = \sum_{g'h'i,...}B^{w,g'h'i'...}_{p,ghi...}(abcdef)\ket{\psi_{abcdef}^{g'h'i'j'k'l'}}. 
\end{equation}
A single such term is then described by 6 index \(F\)-symbols associated to each vertex transformation,

\begin{eqnarray}
\label{StringNet_FSymbols}
B^{w,g'h'i'...}_{p,ghi...}(abcdef)
= F^{al^*g}_{w^* g' l'^*}F^{bg^*h}_{w^*h'g'^*}F^{ch^*i}_{w^*i'h'^*} \nonumber
\\ \times F^{di^*j}_{w^*j'i'^*}F^{ej^*k}_{w^*k' j'^*}F^{fk^*l}_{w^*l'k'^*}.
\end{eqnarray}

Each F-symbol obeys a series of constraints and defines the rules of allowed link combinations \cite{Levin_2005}. It should be noted that explicit partition functions and spectra of these models have been studied in \cite{Vidal_2022,ritzzwilling2023topological}, which must ultimately play into our extensions.

The operators \(B_p^w\) and \(A_s\) commute by construction, supplementing Eq. (\ref{SPPMConstra}). Constraints on \(B_p^w\) operators, analagous to Eq. \ref{SPPMConstrb}, are then realized through the underlying group structure of \(B_p^w\). A global constraint like Eq. \ref{SPPMConstrd} is, again, established for \(B_p^w\) by the aforementioned 
"fat lattice" construction. We can map the \(B_p^w\) operators into site-local degrees of freedom representing the same group, such as we had in the SPPM for classical clocks. Extending these types of constraints to string-net \(A_s\) terms or some decomposition of some \(A^u_s\), similar to \(B_p^w\), would allow us to directly relate many string-nets to classical models. We could then finally establish constraints on bilinear strings of \(A^u_s\) and \(B_p^w\) operators of the same type as Eq. \ref{SPPMConstrc}. Quantum double models are a simple example of this, in the sense of their mapping to both string-nets and classical models. Further difficulties arise when F-symbols introduce complex phases, such as the \(\mathbb{Z}_2\) Chern-Simons phase in \cite{Levin_2005}. In this, plaquettes also include complex phase factors on the legs which can be factored out of \(B^w_p\). Remaining terms are mapped to classical degrees of freedom.  The TQFT then used as the low energy effective theory is a \(U(1) \times U(1)\) Chern-Simons theory. Additional coupling, just as in the SPPM, can change the nature of lowest--energy excitations without changing the ground-state space.

\section{Higher-dimensional models} 
Higher-dimensional systems 
dual to clock models can be constructed from extensions of known stabilizer models. 
For instance, we may map an SPPM type extension of the cubic lattice  \(\mathbb{Z}_q\) TC model \cite{Nussinov_2008_3DTC, Araujo_de_Resende_2019, Kong_2020}
to a classical clock model on the face-centered cubic lattice (FCC) as shown 
in Fig. \ref{3DStarPlaq}. 
This model obeys Eqs. \ref{SPPMConstra}-\ref{SPPMConstrd}. We define it by Eq. (\ref{SPPMHamiltonian}) yet now with six spin star \(A_s = \prod_{j \in s}X_j\) 
and four spin plaquette \(B_p = Z_1Z_2^\dagger Z_3Z_4^\dagger\) operators 
(with indexing, once again, performed  anti-clockwise) as depicted in Fig. \ref{3DStarPlaq}. The local constraint \(\prod_{p \in Cube} B_p = 1\) can be implemented in the classical spin algebra with local defect terms, giving severe further transformations on the Hamiltonian of the FCC spin model.
In this variant of Eq. (\ref{SPPMHamiltonian}) 
the nearest-neighbor condition \(\langle s,p \rangle\) is defined on the 3D cubic lattice. 
We map each face \(B_p\) to a \(z^*_j\) at the center of an FCC face. The star operators $A_s$ at the cubic lattice sites are mapped to \(z_j\) lying on FCC cube corners (Fig  \ref{3DStarPlaq}). 
 Cubic lattice 
 TC models display TQOs which are membrane extensions of string order seen in the (2+1)D case \cite{Araujo_de_Resende_2019, Nussinov_2008_3DTC,Kong_2020}. 

 \section{Brief remarks concerning comparisons with several other models}

 Our construct in which the underlying microscopic theory features effective charges  with geometrically dependent interactions is a very general one. Various known models share some similarities and important differences with the theories belonging to our general framework. We briefly comment on some of these below.

 Rich space filling crystalline structures somewhat similar to those of the anyons that we discussed in the lowest excited states of the SPPM are known to appear in myriad systems including, e.g., the Quantum Dimer Model (QDM) \cite{comp1} 
 that features complex crystalline structures (including devil staircases). The field theories describing the QDM whose derivation relies on Wess-Zumino-Witten type contributions are of an extended Maxwell form \cite{comp2} and, in three dimensions, exhibit higher order curl terms. These field theories enable an understanding of the complex QDM phase diagram (including the said rich crystalline structures). However, when these crystalline structures of dimer coverings  arise, the systems are no longer topologically ordered (unlike the topologically ordered systems that form the focus in our work). 

Along other lines, there are numerous works, e.g., \cite{comp3} 
studying effective TQFT type (including BF) descriptions of systems in which the lowest energy excited states are compact (unlike the SPPM when $8J>g_{s} >0$). These effective descriptions of the lowest lying excited states do not give rise to the divergent number of anyons that, as we established, may arise. 


Given the link between symmetries and topological order \cite{nussinov_ortiz_2009_IsingChains,PNAS-top,generalized,Gaiotto:2014kfa_GenSymmetries}, 
theories may feature topological order when the  underlying higher symmetries of a bare theory are preserved by additional terms \cite{cordova2022neutrino,cordova2022symmbreaking}. 
Our non-TQFT type theories preserve all $d \ge 1  $-dimensional Gauge like symmetries \cite{nussinov_ortiz_2009_IsingChains,PNAS-top} of the simpler (TC and other) models. 
Specific physical implications of such considerations have been suggested in \cite{cordova2022neutrino,cordova2022symmbreaking}  where 1-form symmetries are preserved by Maxwell-Euler-Heisenberg and local operator deformations on the theory. Contrary to our framework, Chern-Simons type modifications may generally break local gauge invariance with very specific quantization schemes enabling gauge invariance 
\cite{dunne1999aspects_chernsimons}. In specific Chern-Simons modifications of the Maxwell theory that preserve gauge invariance, the Chern-Simons term gives rise to discrete topological vacuum states \cite{S-Dual_MCSTheory_Armoni}.  Our construct in which all terms are inherently topological is a rather general method allowing for various interactions and is different from these and other earlier studied theories.  

\end{document}